\documentclass[preprint]{aastex}			
\usepackage{natbib}
\usepackage{amsfonts}
\usepackage{amssymb}
\usepackage{apjfonts}	

\slugcomment{To Appear in the Astronomical Journal}
\shorttitle{VLT $3.8\micron$~Trapezium}
\shortauthors{Lada et al.}
\newcommand{\solarmass}{\ensuremath{ \mathnormal{M}_{\Sun} }}

\newcommand{\av}{\ensuremath{ {A}_{V} }}

\newcommand{\Lp}{\ensuremath{ {L}^{'} }}

\begin{document}
\title{
Deep 3.8\micron~Observations of the Trapezium Cluster\footnote{
Based on observations collected at the European Southern Observatory,
Chile, ESO Program 70.C-0471(A)}}

\author{Charles J. Lada and August A. Muench} 
\affil{Harvard-Smithsonian Center for Astrophysics} 
\affil{Cambridge, MA 02138} 
\email{clada@cfa.harvard.edu, gmuench@cfa.harvard.edu}
\author{Elizabeth A. Lada} 
\affil{Department of Astronomy, University of Florida} 
\affil{Gainesville, FL 32611} 
\email{lada@astro.ufl.edu} 
\and
\author{Jo\~ao F. Alves}
\affil{European Southern Observatory} 
\affil{Karl-Schwartzschild-Strasse 2, 574 Garching Germany} 
\email{jalves@eso.org} 
%
%
%
\begin{abstract}

We present deep 3.8 $\mu$m $\Lp$ imaging observations of the Trapezium cluster in Orion
obtained with the ESO VLT.  We use these observations to:  1) search for infrared
excess emission and evidence for protoplanetary disks associated with the faint,
substellar population of this young cluster, and 2) investigate the nature and
extent of a recently discovered population of deeply embedded sources located in dense
molecular gas behind the cluster.  We detected 38 $\Lp$ sources with substellar
luminosities.  In addition, we detected 24 $\Lp$ sources that were spectroscopically
classified as substellar objects in previous studies.  Examining the infrared colors of
all these sources we determine an infrared excess fraction of 50 $\pm$ 20 \% from the
$JHK_s\Lp$ colors for both the luminosity selected and spectroscopically selected
substellar samples.  This finding confirms the presence of infrared excess, likely due
to circumstellar disks, around a significant fraction of the cluster's substellar
population, consistent with the indications of earlier observations obtained at shorter
($JHK_s$) wavelengths.  Our deep $\Lp$ imaging survey also provides new information
concerning the deeply embedded population of young objects located in the molecular
cloud behind the cluster and revealed in an earlier $L$ band imaging survey of the
region.  In particular, our present $\Lp$ survey doubles the number of sources in the
cluster region known to possess extremely red $K-L$ colors.  These objects exhibit
$K-\Lp$ colors indicative of deeply buried, possibly protostellar, objects that likely
mark the site of the most recent and ongoing star formation in the region.  We find the
surface density distribution of the deeply embedded population to follow that of the
background molecular ridge and to be highly structured, consisting of a string of at
least 5 significant subclusters.  These subclusters may represent the primordial
building blocks out of which the cluster was and perhaps still is being assembled.
These observations may thus provide insights into the early stages of cluster
formation and appear consistent with recent simulations that suggest that
the Trapezium cluster may have formed from numerous but small primordial subclusters. 

\end{abstract}
%
%

\section{Introduction}
\label{sec:intro}

The Trapezium cluster is an extremely young and rich embedded cluster at the heart
of the Great Orion Nebula.  Discovered in early infrared images by Trumpler (1931)
and Baade and Minkowski (1937), it is the best studied of all embedded
clusters.  With age of approximately 10$^6$ years it remains one of the most
important laboratories for investigations of star formation and early stellar
evolution in our galaxy.  In particular, the cluster has a well determined and well
sampled IMF ranging from O stars to substellar objects near the dueterium burning
limit (e.g., Muench et al.  2000, 2002; Lucas \& Roche 2000; Luhman et al.  2000,
Hillenbrand \& Carpenter 2000).  This enables the statistically significant
examination of a number of important astrophysical questions, such as the origin
and nature of the IMF, the frequencies of circumstellar disks and protostars in a
young population, and the origin of stellar multiplicity in the earliest stages of
stellar evolution.  Infrared imaging observations of embedded clusters are
particularly well suited for addressing such issues (Lada \& Lada 2003).  In a
previous infrared imaging study Lada et al.  (2000) obtained extensive {\sl JHKL}
photometry of the cluster.  These observations were used both to measure the
frequency of infrared excess emission in order to constrain the circumstellar disk
fraction in the cluster population and to identify heavily buried protostellar
candidates in this young region.  They found a relatively high excess fraction of
$\sim$ 80\% for the cluster membership, indicating that the vast majority of stars
in the cluster were born surrounded by protoplanetary disks.  Their observations
also revealed a new population of heavily reddened objects deeply buried in
molecular gas behind the main cluster.  These sources are probably extremely young
and mark the location of the most recent and active star formation in the Orion
Nebula region.  An appreciable fraction of these sources could be protostellar
objects. 

For cluster members for which spectra were available, Lada et al.  (2000) found that
the high excess fraction at L-band (3.5 $\mu$m) was independent of spectral type and
thus stellar mass, for stars later than type A.  Indeed, the excess/disk fraction
remained high down to the lowest mass objects near the hydrogen burning limit (HBL).
This raised the interesting question of whether substellar objects were also born with
circumstellar disks.  The answer to this question could provide important clues
concerning the nature and formation of brown dwarfs (Muench et al.  2001, Reipurth \&
Clarke 2002).  Although the $L$-band observations of Lada et al.  were not sensitive
enough to examine a statistically significant sample of objects with substellar
luminosities, they did indicate that 8 of 10 objects with spectral types later than M6
exhibited strong $L$-band excess.  Since M6 (e.g., Luhman et al.  1998) is the expected
spectral type of a young PMS star at the HBL, this result suggested the possibility
that both stellar and substellar objects form surrounded by dusty disks.  Indeed,
examining deeper $JHK$ observations of the Trapezium, Muench et al.  (2001) found
$K_s$-band (2.16 $\mu$m) excesses around a significant fraction ($\sim$ 65\%) of the cluster
members with substellar luminosities.  The fact that many of these brown dwarf
candidates with K-band excesses were spatially coincident with optically identified
{\it Hubble Space Telescope} "proplyds" (e.g., Bally, O'Dell \& McCaughrean 2000)
indicated that these substellar candidates were indeed surrounded by disks.  The
excess/disk fraction implied by the Muench et al.  data was somewhat lower than that
derived for the stellar objects in the cluster from the $JHKL$ band observations, but
higher than predicted by disk models for brown dwarfs which produce 
smaller amounts of excess emission at $K$-band (e.g., Walker et al.  2004; Natta et al.
2002)).  Moreover, a number of the brown dwarf excess sources were found to lie well
below the expected disk locus on the $JHK$ color-color diagram, also inconsistent with
expectations of circumstellar disk models.
Because the magnitude of the infrared excess from a disk increases with wavelength and
only very small amounts of dust are required to produce optically thick emission in the
3 - 5$\mu$m range, $L$ band is the optimum wavelength for detecting infrared excesses
from circumstellar disks using ground-based telescopes (Haisch, Lada \& Lada 2001, Wood
et al.  2002).  To better constrain the excess/disk fraction for the substellar
population of the Trapezium cluster requires an extension of the previous L band
imaging survey into the brown dwarf regime.  Deeper $L$ band observations would also be
capable of improving our knowledge and understanding of the deeply embedded population
behind the cluster.  Do these objects represent a continuation of the cluster forming
process that produced the more revealed Trapezium cluster?  If so, then their
observation may provide interesting insights concerning the nature of the physical
mechanism of cluster formation.  For example, Scally and Clarke (2002) have argued that
the Trapezium cluster could have been formed from the merger of numerous small
primordial subclusters.  In this case we might expect to directly detect such
subclustering in the embedded population.  Observation of structure or the lack of it
in the embedded population can thus provide important constraints on models of cluster
formation.  To obtain a sufficiently deep 3 $\mu$m survey of the Trapezium cluster
requires a large telescope, a sensitive array of $L$ band detectors and the best seeing
possible.  To achieve these goals we obtained observations of a significant portion of
the Trapezium cluster with the ESO VLT using an $\Lp$ (3.78$\mu$m) filter under
conditions of good seeing.  In this paper we present these data, their analysis and
implications for understanding the nature of extremely young brown dwarfs and the early
stages of cluster formation.

%
%
\section{Observations and Data Reduction}
\label{sec:images:obs}



We obtained deep imaging of the Trapezium cluster in the $L^\prime$ band at 3.78 $\mu$m
using the Infrared Spectrometer And Array Camera, ISAAC (Moorwood et al.  1998), at the
VLT UT1 telescope.  In order to obtain the best seeing conditions at Paranal the
observing program was carried out in Service Mode and data acquisition took place
during several nights of good seeing during the southern Summer of 2002.  The
$L^\prime$-band sky coverage is shown in Figure 1. Table 1 is a log of the
observations and lists the positions, integration times and the measured seeing
sizes (FWHM) for each of the fields.


To remove the sky at the $L^\prime$-band we observed in chopping mode using the VLT
chopping secondary mirror in phase with the detector readout.  To obtain a difference
image the telescope was also nodded 20 arcsec in right ascension in a standard ABBA
pattern.  The chopping throw was 20 arcsec in right ascension at a frequency of 0.43
s$^{-1}$.  Each on and off image consisted of 9 coadded frames with an integration time
of 0.11 seconds per frame.  Each nod consisted of 15 chop (on-off) cycles.  Each
observation consisted of 15 ABBA nod cycles with a random jitter of up to 20 arcsec
between ABBA sets.  This resulted in a total integration of typically 29.7 minutes for
stars observed in both the on and off chop beams.  This observing technique gives the
best sky subtraction in the $L^\prime$-band, although it generates a final image that
contains many negative sources resulting from sky beam counterparts at the chopping
distances.  Despite this very few known sources (15) fell into negative images in the
reduced chopped image.

Because the chop throw was smaller than an individual image, most stars appeared in
both the on and off images and were thus observed twice in each cycle.  To take
advantage of this each member of the on-off cycle was sky subtracted using the other
member of the cycle as a sky to produce two positive images of a target star for each
chopped cycle.  These images were then flat fielded using sky flats.  The flattened
images were then appropriately shifted and stacked to produce the final combined image,
which typically consisted of a total of 120 observations of each star.  This procedure
resulted in the production of a final image that was rectangular ($\sim$ 135 x 85
arcsec) in shape.  Some of our final combined images contained only 118 observations
since occasional problems with the chopping secondary resulted in the production
elongated images on some frames and the few observations with such poor images were
discarded before the final image was constructed.  Figure 2 displays the combined
chopped $\Lp$ band image for one of our fields (L5) and illustrates the quality of the
final images.  The seeing was typically below 0.5 arcsec.


Sources were identified by visual inspection of the VLT chopped images 
and by comparison to the locations of sources known from published 
near-infrared catalogs (e.g., Muench et al 2002, Hillenbrand \& Carpenter 2000).
Our final catalog consisted of 424 sources with measurable
$\Lp$ photometry; indeed, only 15 known sources were not detected due to negative 
chop  images while 7 bright sources were saturated.  By comparison to 
ground-based $JHK$ surveys we find that we detected $\sim98\%$ of the 
known sources brighter than $H=16$ (see also Figure \ref{fig:vlt_cmds}b).  
Further, 15 new $\Lp$-only  sources were identified with the vast 
majority of them projected along the OMC-1 cloud core. Last, $\sim100$ 
individual sources appeared on two or more VLT images (see Figure \ref{fig:area}), 
yielding quality checks on the photometric and positional accuracy of our final catalog.

The IRAF APPPHOT package was used to obtain aperture photometry at a 
large range of radii ($2-12$ pixels) and with a sky annulus from $12-22$ pixels.
Corresponding aperture corrections were calculated using the MKAPFILE 
routine and 10-20 bright sources for each VLT image, although the images' 
relatively similar PSFs yielded similar corrections. 
For this reason, we chose the 8 pixel aperture photometry for all sources,
corresponding to a beamsize of  $1.136\arcsec$ and a typical aperture 
correction of $\sim-0.13$ magnitudes.  
The 8 pxiel photometry of 13 sources were affected by negative chop 
images and/or bright stars in the wings of their PSFs; their photometry 
was adjusted to an appropriately smaller beamsize.
Zeropoint calibration was performed using standard star observations obtained
on 3 of the 4 queue observing nights, and photometric zeropoint offsets between
frames as revealed by sources in overlap images were removed by cross-calibration 
to the photometry of the L1 position.
Our quoted photometric accuracy as derived from overlap sources is $\sim5\%$.

Sources were placed onto the 2MASS astrometric grid using plate solutions derived with
IRAF routine CCMAP from 30-60 sources on each reduced image.  Formal rms errors of
these solutions were $\sim0.1\arcsec$; however, sources in overlap regions reveal a
smaller relative scatter of order $0.05\arcsec$.  The final catalog is available in
electronic format with Table \ref{tab:catalog} representative of its
contents.  To preserve positional information lost in the astrometric plate solutions,
we provide both equatorial coordinates and pixel positions of each source (on a
specific VLT image).  Sources are also cross-referenced to the FLWO-NTT (Muench et al
2002), Hillenbrand \& Carpenter (2000) and Hillenbrand (1997) identifications.

%
%

\section{Results and Discussion}
\label{sec:results}

\subsection{Color-Magnitude Diagrams: Defining the Substellar Sample}
\label{sec:results:CMDs}

In order to describe the population of $\Lp$ sources in our survey region,
illustrate the sensitivity of our VLT observations, and select the candidate
substellar population in the cluster we present the $K-\Lp$ vs $K$ and $H-K$ vs $H$
color-magnitude diagrams (CMDs) in Figure \ref{fig:vlt_cmds}.  A total of 400
sources are detected and resolved in $K$ and $\Lp$ passbands; these are plotted in
Figure \ref{fig:vlt_cmds}a.  Most sources have $K-\Lp\,<\,2.0$; however, a
significant population of sources have much larger colors, with a handful of
sources characterized by extremely red colors (i.e., 5 $\leq$ $K-\Lp$ $\leq$ 6
magnitudes) suggesting reddenings up to $\av\,\sim\,100$.  It is unclear if the
inferred reddenings to these extreme sources are due to pure dust extinction or
strong infrared excess emission of the kind that is often associated with
protostellar objects, since the reddest of these are not detected at $J$ or $H$
bands.

Similar to Lada et al.  (2000), we define the depth of our survey by our ability to
detect the photospheres of cluster members.  In Figure \ref{fig:vlt_cmds}b we plot the
$H-K$ vs $H$ color-magnitude diagram for all sources known from previous deep IR
surveys within the boundaries of the VLT $\Lp$ survey.  Presuming that the $H$
magnitude of these sources is primarily photospheric, we clearly detect the vast
majority of sources down to $H\,\sim\,16$, below which our observations are quite
incomplete.  Alternately, we are sensitive to the photosphere of a 1 Myr brown dwarf
($0.075\solarmass; H_0\,=\,13.54$; Baraffe et al 1998) seen through 20 visual
magnitudes of extinction.

Muench et al.  (2001) defined a sample of Trapezium sources that were postulated to
be sub-stellar based on their unreddened $H$ band luminosity from the 1 Myr
isochrone of Baraffe et al.  (1998).  The validity of that CMD analysis was
bolstered by the fact that the luminosity selected sources were on the whole
fainter than members with spectral types earlier than M6.  Following Muench et al.
(2001), we use the $H$ vs $H-K$ color-magnitude diagram to identify candidate brown
dwarfs in the Trapezium cluster by comparing their infrared luminosities to the
predictions of the Baraffe et al.  models.  We consider all sources whose
dereddened luminosities are less than the the predicted luminosity of the HBL for
the adopted age (10$^6$ yrs) and distance (400 pc) of the cluster.  By considering
all sources below the the extrapolated reddening vector for the HBL we identify 38
sources as substellar candidates in our VLT fields.  However we note that this
sample represents an upper limit to the true number of substellar cluster members
that we detected at $\Lp$ since we have not accounted for background/foreground
field star contamination which is predicted to be significant at these faint
magnitudes.  For example, Muench et al.  (2002) estimated, by combining
observations of control fields near but off the cluster and a model of the extinction
distribution produced by the molecular cloud, that field stars account for as much
as 20-30 \% of the objects in this magnitude range.


\subsection{Color-Color Diagrams: Measuring the Infrared Excess Fraction}
\label{sec:results:cc}

In earlier studies Lada et al (2000) and Muench et al (2001) used $JHKL$ and
$JHK_s$ color-color (CC) diagrams, respectively, to diagnosis the presence of
infrared excess due to disks around stellar and candidate substellar members of the
Trapezium Cluster.  We now extend their analysis to include high dynamic range VLT
$\Lp$ data to better assess the infrared excess fraction across the HBL and into
the substellar regime.  In Figure \ref{fig:vlt_jhkl} we present the $JHK_s$ and
$JHK_s\Lp$ infrared CC diagrams for the VLT $\Lp$ Trapezium sources.  
There are a total of 345 sources with $JHK\Lp$ photometry. Sources represented by 
filled circles are the 38 luminosity selected substellar candidates from the 
$H$ vs $H-K$ color-magnitude diagram. Plotted for comparison
is the intrinsic locus for dwarfs taken from a merger of optical and IR colors from
Winkler (1997), Kenyon \& Hartmann (1995), Bessell \& Brett (1988) and Leggett
(1992).  Reddening vectors (Whittet 1989) for the giant branch and M6 and M9
dwarfs are also shown.

Although many sources display colors consistent with reddened photospheric colors, a
large fraction (50\% and 67\%, respectively) display infrared excess relative to the M6
reddening vector on both the $JHK$ and $JHK\Lp$ CC diagrams.  As expected, the excess
fraction is found to be higher in the latter CC diagram.  For the candidate 10$^6$ year
old substellar population we expect most objects to have spectral types of M6 or later
(Luhman et al.  1998).  Since most of the stars plotted in Figure \ref{fig:vlt_jhkl}
are more luminous than the HBL, and thus likely to be earlier than M6, those that
appear in the infrared excess region of the CC diagrams (Lada \& Adams 1992) must be
characterized by significant $H-K$ and $K-\Lp$ excesses.  Identifying infrared excess
for substellar objects, particularly in the $JHK\Lp$ CC diagram, is a much more
difficult task than for it is for stars above the HBL.  This is because the range of
intrinsic $K-\Lp$ colors for M6-M9 dwarfs is relatively large and, in fact, comparable
to the entire range in intrinsic $K -\Lp$ colors for all earlier spectral types and
moreover larger than the corresponding $H-K$ intrinsic color spectrum.  Nevertheless, a
significant fraction (42\%) of the substellar candidates fall to the right of the M9
reddening vector on the $JHK\Lp$ CC diagram and consequently have significant $K-\Lp$
excess.  A similarly high fraction (44\%) of substellar candidates display excesses in
the $JHK$ CC diagram with respect to the M9 reddening vector.  This relatively high
fraction from the $JHK$ CC diagram is roughly consistent with the findings of Muench et
al.  (2001) but inconsistent with expectations of disk models, as mentioned earlier.

We also note that about 1/3 of the luminosity selected substellar objects have colors
apparently consistent with reddened stars of spectral type earlier than M6.  Some of
these sources are likely field stars.  For others their relatively blue colors could be
the result of photometric uncertainties in the observed colors.  Moreover, the $JHK$
observations were obtained at a different epoch than the $\Lp$ observations, and
consequently variability, if present, could contaminate some of the $K-\Lp$ colors.  In
addition, this result is also reminiscent of the recent spectroscopic study of
Slesnick, Hillenbrand and Carpenter (2004) who found a similar sized fraction of
luminosity selected substellar candidates in the Trapezium cluster to be characterized
by spectral types earlier than the nominal M6 substellar boundary.  The nature of this
apparently subluminous stellar population is unknown.  Both our CC diagrams also show
sources which fall into relatively 'forbidden' regions of the diagrams.  In Figure
\ref{fig:vlt_jhkl}a there is a population of sources with very red $H-K$ colors and
relatively blue $J-H$ colors.  Some of these same sources also appear to be
unexpectedly blue in $K-\Lp$ color in Figure \ref{fig:vlt_jhkl}b; as was discussed in
Lada et al (2000) and Muench et al (2001) many of these sources are 'proplyds' (O'dell
Wen \& Hu 1993) and their colors are possibly contaminated by scattered light and/or
emission from the photoionized envelopes surrounding the underlying disks.


To further investigate the behavior of the infrared excess/disk fraction across the HBL we
divided the source distribution in the $H-K$ vs $H$ CMD into 7 luminosity bands,
approximating equal sized steps in unreddened $H$ magnitude and logarithmic mass using
predicted $H_0$ magnitudes from the the Baraffe et al (1998) 1 Myr isochrone.  The magnitude
bins and corresponding predicted masses are listed in Table \ref{tab:trap_cc}.  We further
constructed an extinction limited sample by including in each bin only sources with
extinctions $\av \leq 20$.  The fraction of sources in each luminosity selected bin
displaying infrared excess greater than their formal 1 sigma photometric errors was counted
in the corresponding $JHK$, $JHK\Lp$ and $HK\Lp$ color-color diagrams.  The M6 spectral type
boundary was used to calculate the infrared excess fraction for all bins.  In addition for
bins 5-7 the infrared excess fraction was also calculated using the M9 boundary, since these
bins are likely to contain brown dwarfs.  The corresponding excess fractions are listed in
Table \ref{tab:trap_cc}, while in Figure \ref{fig:trap_cc_irx} we plot these fractions as a
function of the luminosity bin center.

Clearly, Trapezium sources display excess at all magnitudes and in all diagrams
with the $JHK\Lp$ and $HK\Lp$ diagrams giving nearly identical results.  As already
shown by Lada et al (2000), the fraction of bright sources displaying excess is
larger when using $K-L$ as a color diagnostic \footnote{The fraction of bright
($H<12$) sources showing excess in these diagrams appears lower than published in
Lada et al (2000) for two reasons.  First, the $\Lp (3.8\micron)$ bandpass of the
VLT data yields redder photospheric colors than the $L (3.5\micron) $ bandpass of
the FLWO data making the identification of a infrared excess somewhat more
difficult at $\Lp$ for late spectral types.  Second, we are consistently counting
from the M6 rather than the M5 boundary used Lada et al (2000)}.  Yet as $H_0$
increases into the fiducial brown dwarf regime (Band 5-7), the excess fraction
traced by $K-\Lp$ appears to decrease, although the uncertainty in the excess
fraction increases due the larger range of intrinsic $K-L$ colors for M6-M9
sources.  Note, our results for band 7 are likely upper limits because of
incompleteness, which manifests itself in our ability to detect only those faint
sources with disks.  This decrease in $K-L$ excess fraction is {\em opposite} the
behavior traced by $H-K$, which increases for the brown dwarf luminosity bins and
recovers the $>50\%$ excess fraction quoted by Muench et al (2001).  The physical
origin of this difference in behavior is unknown.  Nonetheless, the $\Lp$
measurements indicate a relatively high infrared excess and likely disk fraction
for the substellar population of this cluster.  Merging the results for luminosity
bands 6 and 7 from Table 2 we estimate an infrared excess fraction for the
substellar population of the cluster of 50 $\pm$ 20 \% from the $JHK\Lp$ diagram.
Our measurements also hint that the disk fraction for the substellar population is
somewhat lower than that of the stellar population.  However, considering the
uncertainties and the possible effects of field star contamination, one cannot yet
draw a firm conclusion about the presence of a decrease in excess fraction for the
substellar population in the cluster.

There are a number of limitations to our estimate of the excess/disk fraction for
the substellar population from our VLT $\Lp$ data.  First, our measurements of the
excess fraction are incomplete for the lowest luminosity bin resulting in an
overestimate for the excess fraction in that bin.  Second, the lack of specific
knowledge about the intrinsic infrared colors of the objects results in the
relatively large quoted uncertainties ($\pm20-30\%$) for our estimate of the excess
fraction of the luminosity selected substellar candidates.  The lack of knowledge
of source intrinsic color also presumably affects the derived excess fraction at
brighter magnitudes as well since an M0 star, for example, will require 0.3
magnitudes of $K-L$ excess in order to be counted as an excess source using the M6
boundary.  Given that any age spread for Trapezium members will result in a mix of
mass (as traced by spectral type) as a function of luminosity, assigning a single
spectral type boundary only acts to underestimate the true excess fraction.  In
addition we have not attempted to correct for foreground/background source
contamination which is likely present for these faint sources.  The presence of
such contaminating sources also results in an underestimate of the true excess
fraction.

 
\subsection{Infrared Excess Fraction for a Spectroscopically
Selected Substellar Sample}
\label{sec:irx:spt}

Liu et al.  (2003) have shown that a more complete assessment of infrared excess in
substellar sources can be made if the spectral types and thus the intrinsic colors of
the stars are individually known and taken into account in measuring infrared excess
emission.  Combining spectra with infrared $JHK\Lp$ photometry of substellar candidates
in Taurus and IC 348, Liu et al.  (2003) derived a $K-\Lp$ band excess/disk fraction of
77\% for stars they selected with late spectral types.  This was a significantly higher
fraction than that ($\sim$ 33\%) inferred from analysis of the $JHK\Lp$ CC diagram for
the same sources.  Therefore it would be extremely valuable to examine the infrared
colors of those substellar candidates in the Trapezium that have been spectroscopically
classified so that their intrinsic infrared colors are known.

Over the past few years various workers have collected spectra of faint stars in the
Trapezium cluster (i.e., Hillenbrand, 1997; Lucas et al.  2001; Luhman et al.  2000;
Slesnick, Hillenbrand and Carpenter, 2004).  From these studies we find a total of 24 objects
with spectral types of M6 or later that were also detected in our $\Lp$ images.  In Figure
\ref{cc_w_spectra} we show the $JHK_s$ and $JHK_s\Lp$ CC diagrams for this spectroscopically
selected sample of substellar candidates.  With the assumption of dwarf colors for the
spectroscopically selected sample we can convert the spectral types to intrinsic substellar
colors.  With knowledge of the spectroscopically derived intrinsic $H-K_s$ colors of the
stars, we derive an excess fraction of 70\% $\pm$ 15\% from comparison with their observed
$H-K_s$ colors.  This is in excellent agreement with the fraction derived by Muench et al.
(2001) from photometry using the $JHK_s$ CC diagram and roughly consistent with our estimate
based solely on $JHK_s$ photometry of sources detected at $\Lp$.  However this value is
higher than would be expected by traditional disk models.  Using the $K_s-\Lp$ data we derive
in a similar manner an excess fraction of 52\% $\pm$ 20 \% for the spectoscopically selected
substellar candidates.  Although this represents a significant excess fraction, the
calculated value is formally lower than both the fraction derived by Liu et al.  (2003) for
Taurus and IC 348 using $K-\Lp$ colors and the fraction derived for the same sources from our
$JHK_s$ colors.  However given the relatively large uncertainties it is difficult to assess
the significance of this difference.  Moreover it is difficult to compare the results between
our spectroscopically selected Trapezium sample and the Liu et al.  $\Lp$ IC 348 plus Taurus
sample, since neither can be considered complete and may not even be fully representative of
the entire substellar populations in the respective clouds.  Finally, an additional
uncertainty may arise from our assumption of dwarf colors for these objects since they are
expected to have lower surface gravities than field dwarfs.  Theoretical models predict that
lower surface gravity stars should have somewhat bluer $K-L$ colors, leading to an
underestimate of the excess fraction when assuming dwarf colors (Liu et al.  2003).  We find
that these same model atmospheres (i.e., Baraffe et al.  2002) predict slightly bluer $J-H$
colors and no change in the $H-K$ colors for lower surface gravity atmospheres with the result
that the excess fraction derived assuming dwarf colors is not significantly altered.  The
relatively large excess fraction derived from $JHK$ colors is unexpected and as speculated
earlier by Muench et al.  (2001) may be the result of scattered light or emission from the
photoionized envelopes of the disks, since many of the sources with these 
colors appear to be proplyds.

\section{Deeply Embedded Objects and Protocluster Structure}

The $L$ band imaging survey of Lada et al (2000) resulted in the identification of a
previously unrecognized population of heavily reddened objects toward the Trapezium cluster.
Seventy-eight sources were found to have large $K-L$ colors (i.e., $K-L$ > 1.5 mag.)
suggestive of objects deeply embedded in molecular gas and dust.  For example, for a normal
interstellar reddening law (e.g, Mathis 1990) this would correspond to $A_V$ $>$ 27
magnitudes for a typical naked field star seen through the cloud and approximately $A_V$ $>$
15 magnitudes for a young star with a circumstellar disk.  The spatial distribution of these
red objects was found to differ from that of the bulk of the (less reddened, $A_V \sim$ 5
mag.)  cluster membership, but to be similar to that of the dense molecular ridge located
behind the cluster.  The molecular ridge has long been known to be a site of star formation,
haboring the massive protostellar objects IRC2 and BN-KL.  The $L$ band survey indicated that
the star formation activity in the ridge was considerably more active than previously
suspected.  Many of the new and heavily buried sources were found to be characterized by the
extreme infrared excess in the $JHKL$ CC diagram often observed in protostars.

Our deeper $\Lp$ observations have revealed 73 objects with $K-\Lp$ $>$ 1.5 magnitudes
and 40 with $K-\Lp$ $>$ 2.0 magnitudes within the six fields we imaged.  Within these
fields 25 objects were identified as having $K-L$ $>$ 1.5 magnitudes in the earlier
survey by Lada et al.  (2000) and 48 are newly identified members of the deeply
embedded population (most of which were below the $L$ band sensitivity limit in the
previous survey).  Together with the earlier data of Lada et al., our observations
indicate that that deeply embedded population consists of approximately 126 members,
roughly 20\% the size of the foreground and more revealed Trapezium cluster.

The VLT sources with $K-\Lp$ $>$ 1.5 magnitudes are plotted as filled squares in Figure 1.
These sources form a tight distribution that more or less follows the molecular ridge (see
figure 7 of Lada et al.  2000 and also Johnstone \& Bally 1999).  The extremely high
extinctions inferrred for these sources coupled with their relatively high surface density,
clearly indicates that the vast majority are embedded in the cloud and not reddened
background field stars.  Moreover, the reddest of these sources have $K-\Lp$ colors
suggestive of visual extinctions of up to 100 magnitudes!  Such extreme extinctions typically
arise from dense envelopes of deeply embedded protostellar objects and not from the more
distributed dense gas of a molecular cloud core.  As suggested by Lada et al (2000) many such
objects are likely protostellar in nature and still in the process of forming.  In addition
our new observations revealed 15 previously undetected $\Lp$~only sources, plotted as filled
diamonds in Figure 1.  As seen in this figure twelve of these are found coincident with the
molecular ridge and are likely additional members of the deeply embedded population.  The
brightest of these is the source vlt-129 ($\Lp\,=\,10.98$) which lies $28\arcsec$ southwest
of the BNKL and is characterized by a $K-\Lp$ color of $>$ 6 magnitudes.

The population of deeply embedded objects (DEOs) traces the most recent epoch of star
formation in the Trapezium cluster region.  The spatial structure of the deeply
embedded population possesses an imprint of the physical processes responsible for its
creation and thus can provide important insights into the earliest phases of cluster
formation process.  In their study of embedded clusters Lada \& Lada (2003) found that
embedded clusters could be characterized by two basic structural types.  Type 1
clusters exhibited hierarchical spatial structure characterized by significant
sub-clustering in their surface density distributions.  Such highly structured systems
are thought to be a signature of the important role of turbulence in the evolution of
molecular gas to form stars.  Type 2 clusters exhibited centrally-concentrated
structure typically characterized by relatively smoothly varying radial surface density
profiles.  Such centrally condensed structures are thought to be a signature of the
global dominance of gravity in the formation of the stellar system.  It is not clear
whether these two types of structures represent two different primordial structures or
two differing evolutionary states of an embedded cluster.  Most embedded clusters
appear to be type 2 structures, yet most progenitor molecular clouds appear to be
filamentary in shape and highly structured, at least on large scales.  This perhaps
suggests that centrally condensed type 2 structures may form from more structured type
1-like initial configurations.  Indeed, in a recent numerical simulation of cluster
formation from a turbulent cloud, Bate, Bonnell and Bromm (2003) found that filamentary
clouds could form stars which fall together to form small groups within a larger
cluster forming system.  Earlier numerical simulations by Scally and Clarke (2000) have
suggested that relaxed, centrally condensed stellar clusters can form from the merger
of smaller subclusters.  They suggested that the Orion Nebula Cluster (and by
implication the Trapezium cluster which forms its inner core) could have formed from
the merger of subclusters provided that these subclusters were relatively small and
numerous.

Existing surface density maps of the Trapezium cluster suggest that it is a type 2 or
centrally concentrated embedded cluster (Hillenbrand \& Hartmann 1998; Lada et al.
2000; Scally \& Clarke 2002).  In Figure \ref{proto_cluster_map} we plot the surface
density distribution of the embedded population (contours) along with that of the less
extincted members of the Trapezium cluster (grey scale).  Our deeper observations allow
us to map these distributions at higher angular resolution than previously possible.
The deeply embedded population is found to be significantly structured appearing to
consist of five subclusters oriented more or less along the filamentary molecular
ridge.  The basic parameters of the subclusters are listed in Table
\ref{tab:subclusters} These five subclusters account for approximately 55\% of the
stars in the deeply embedded population.  If the formation of the deeply embedded
population is a continuation of the same physical process that produced the foreground
Trapezium cluster, then it is tempting to identify the subclusters with the fundamental
units or building blocks out of which the cluster continues to be assembled.  If this
interpretation is correct, then it suggests that a centrally condensed cluster may be
formed from the merger of smaller subclusters which themselves formed intitially in
filamentary dense molecular gas in accord with the suggestion of Scally and Clarke
(2002).

The deep infrared observations also allow us to examine the structure of the foreground
Trapezuim cluster in more detail than previously possible.  Infrared $JHK$ CC diagrams
(see Figure 4 and Lada et al.  2000, Figure 8) show that the bulk of the Trapezium
cluster is characterized by extinctions less than about 5-6 magnitudes.  Therefore to
map the structure of the foreground Trapezium cluster we ploted in Figure
\ref{proto_cluster_map} the spatial surface density distribution of stars with $J-H <$
1.5 magnitudes (corresponding to $A_V$s of approximately 10 magnitudes or less).  The
surface density map of the cluster is dominated by a strong central peak at the site of
the Trapezium (N$_*$ $=$ 6000 pc$^{-2}$), in agreement with previous studies.  However,
our higher resolution map reveals additional significant structure in the cluster.
Three secondary peaks with peak surface densities of 3000 pc$^{-2}$ or greater are
found to the north and east of the primary peak.  Together the four subclusters account
for about 28\% of the total population of the foreground cluster.  The basic parameters
of these subclusters are also listed in Table \ref{tab:subclusters}.  These secondary
peaks are reminiscent of the satellite subclusters observed in IC 348 (Lada \& Lada
1995) another otherwise highly centrally concentrated embedded cluster.

Comparison of the embedded source and cluster surface density distributions generally
confirms earlier results in that, for the most part, the two populations have different
spatial distributions which are offset from one another.  However, one peak in the DEO
distribution is coincident with the primary peak in the Trapezium cluster distribution.
Although significant this peak in the DEO distribution may not be in fact as prominent
as it appears in the map since at least 3 of the sources in it are foreground proplyds
with very red colors.  In addition the strongest peak in the DEO distribution, which is
coincident with BN-KL region, is near, but offset from the second strongest peak in the
foreground cluster distribution.  Thus it appears that at least some of the structure
in the foreground Trapezium cluster reflects the more primordial structure in the
background embedded population.  These observations suggest that the foreground
Trapezium cluster is still sufficiently young that it is not yet fully relaxed into a
centrally condensed, isothermal-like stellar distribution.  Moreover, the presence of
structure in the embedded source population suggests that the cluster will need to
undergo still more structural evolution in order to eventually incoporate the presently
embedded sub-clusters and form a fully relaxed and centrally concentrated stellar
system.

\section{Summary and Conclusions}

We have obtained deep 3.8 $\mu$~m $\Lp$ observations of the Trapezium cluster using
the ESO VLT.  We have imaged a significant fraction of the inner 5' x 5 ' region of
the cluster.  Our observations extend the previous $L$ band survey of Lada et al.
(2000) to fainter cluster members and include approximately 38 objects with
luminosities below that expected for cluster members at the HBL.  In addition, we
detected 24 objects with previously known spectral types that are later than that
corresponding to the HBL (M6).  Combining our deep $\Lp$ data with previous $JHK_S$
observations we have examined the frequency of infrared excess emission among the
candidate substellar population.  We derive an infrared fraction of 50 $\pm$ 20\%
for the substellar population from analysis of the $JHK_s\Lp$ colors of both
luminosity selected and spectroscopically selected samples of substellar
candidates.  This result suggests a significant disk fraction for substellar
sources, consistent with earlier results for this and other clusters (Muench et al.
2001; Liu et al.  2003).  The presence of disks around substellar objects provides
evidence that the formation mechanism for brown dwarfs and more massive stars is a
physically similar process as has been previously argued (Muench et al.  2001, Liu
et al.  2003).  Due to issues of completeness and contamination of the sample by
foreground/background stars we cannot determine the true excess/disk fraction for
the substellar population and it is therefore not clear whether our measured value
of $\sim$ 50\% for substellar objects is significantly different from the $\sim$
65\% fraction found for a complete and representative sample of the stellar members
of the cluster from $JHK_s\Lp$ observations.

Our observations also confirm that a number of substellar candidate sources possess
unusual colors on the $JHK_s$ CC diagram.  This is true for both luminosity
selected and spectroscopically selected substellar candidates.  These sources
exhibit significantly larger infrared $H-K_s$ excesses than predicted by
conventional disk models and thus appear redder for their $J-H$ color.  Moreover, a
few of these sources exhibit $K_s-\Lp$ colors that are two blue for their $J-H$
color and do not appear as $K_s-\Lp$ excess sources on the $JHK_s\Lp$ CC diagram.
The origin of these anomalous colors is unclear.  Photometric errors, source
variability and perhaps contamination by proplyd emission may all contribute to
explain these unusal characteristics.

Our deep $\Lp$ observations also reveal a significant number of new members of the
deeply embedded population of young objects which lies buried in the molecular cloud
behind the cluster.  As a whole this embedded population represents the location of the
most recent star forming events in the Trapezium region.  We find the surface density
distribution of the deeply embedded population to follow that of the background
molecular ridge and to be highly structured, consisting of a string of at least 5
significant sub-clusters.  If the embedded population represents a continuation of the
star formation process responsible for the creation of the more revealed Trapezium
cluster, then these embedded sub-clusters are likely representative of the primordial
building blocks out of which the cluster was and perhaps is still being assembled.
Additional evidence for this possibility is found in high resolution map of the surface
density distribution of the Trapezium cluster itself which shows, in addition to a
strong and centrally concentrated peak, significant substructure in the form of small
satellite subclusters similar to those in the background cloud. These findings are
generally consistent with the suggestion of Scally and Clarke (2002) that the 
Trapezium cluster could have been assembled from the merger a large number of 
small ($<$ 50 stellar members) subclusters over the past 10$^6$ years. 

\acknowledgments

We are grateful to Richard J. Elston for useful advice and suggestions
pertaining to the image reduction and analysis and to Kevin Luhman
for useful discussions. EAL acknowledges support from 
NSF grants, AST97-3367 and AST02-02976 to the University of Florida.

\clearpage

%
%

\clearpage

%
\clearpage
%
%
\figcaption[fig01.eps]{   
    Spatial distribution of VLT detected sources in the Trapezium. The area covered by individual chopped
    pointings are outlined and labeled L1...L6. Known sources (from the Muench et al 2002 catalog) are
    plotted as open symbols, while those detected and photometered in the VLT $\Lp$~data are shown as filled
    circles.  Open circles containing crosses were either saturated or affected by negative images in the
    chopped data. Large grey filled diamonds mark the locations of 15 new sources detected only in the
    VLT $\Lp$~band data.
\label{fig:area}}

\figcaption[fig02.eps]{   
    VLT $\Lp$ band image of the L5 pointing in greyscale using a logarithmic stretch. Sources detected only
    at $3.8\micron$ are surrounded by white squares. Note the excellent image quality and high spatial 
    resolution, which allows excellent photometry even in the crowded region near the 
    central OB stars (image left).
\label{fig:image}}


\figcaption[fig03.eps]{   
    Infrared color-magnitude diagrams for sources in the VLT Trapezium $\Lp$~images. 
    A) $K-\Lp$ vs $K$; B) $H-K$ vs $H$.
    The location of all sources detected in $K$ and $\Lp$ bands
    are plotted in (A). Sources currently undetected at $H$ band are plotted 
    as open circles; Saturated sources were excluded.
    The locations for all known sources detected in the $J, H$ and $K$ bands
    in the fields observed     by the VLT are plotted in (B), excluding sources that 
    fell into chopped negative regions. Sources detected at $\Lp$ are shown as filled circles. 
    The source distribution in this CMD was divided into 7 bands based on the 
    sources' unreddened
    $H$~band magnitudes;  each luminosity band approximates equal sized steps in logarithmic 
    mass using the 1 Myr isochrone from Baraffe et al. (1998) and a distance of 400 pc. 
    See text. In addition reddening vectors with lengths of  $\av\,=\,$ 35 and 20 
    magnitudes are plotted in A and B, respectively.
\label{fig:vlt_cmds}}

\figcaption[fig04.eps]{   
   Infrared color-color diagrams for sources in the VLT Trapezium $\Lp$~images. 
   A) $J-H$ vs $H-K$; B) $J-H$ vs $K-\Lp$.  Substellar candidate sources
   selected from the CM diagram are indicated by open circles.
   Colors are compared to the dwarf star locus (see text references).
\label{fig:vlt_jhkl}}


\figcaption[fig05.eps]{     
    Fraction of sources displaying excess in color-color diagrams as a function of 
    un-reddened $H$ band magnitude.  Excess fractions were calculated for the luminosity
    bands defined in Figure 3b. 
    Results for two color-color diagrams ($JHK\Lp$ and $JHK$) and 
    two intrinsic color boundaries (M6 \& M9) are shown.  The last bin likely represents an
    upper limit due to incomplete detections at $\Lp$.
\label{fig:trap_cc_irx}}

\figcaption[fig06.eps]{     %
    Infrared color-color diagrams for spectroscopically selected substellar
    candidate sources detected at $\Lp$. Different symbols correspond to
    the different spectral types.
\label{cc_w_spectra}}

\figcaption[fig07.eps]{     
    Maps of the surface density distribution of the deeply embedded object population
    (red contours) and the more foreground and less extincted Trapezium cluster
    (grey scale). For the cluster only members with $J-H$ colors $<$ 1.5 magnitudes 
    (roughly corresponding to $A_V \leq$ 8-10 magnitudes) are plotted. For the 
    deeply embedded population all sources with $K-L$ colors $>$ 1.5 magnitudes 
    (roughly corresponding to $A_V$ $>$ 15-26 magnitudes) are plotted. Contours 
    that trace the deeply embedded population start
    at a surface density of 1000 stars pc$^{-2}$ and increase in steps of 500
    stars pc$^{-2}$. The grey scale contours that trace the cluster 
    start at 500 stars pc$^{-2}$ and increase in steps of 500 stars pc$^{-2}$.
\label{proto_cluster_map}}

%

\clearpage
\newpage
\begin{center}
\includegraphics[angle=90,totalheight=4.5in]{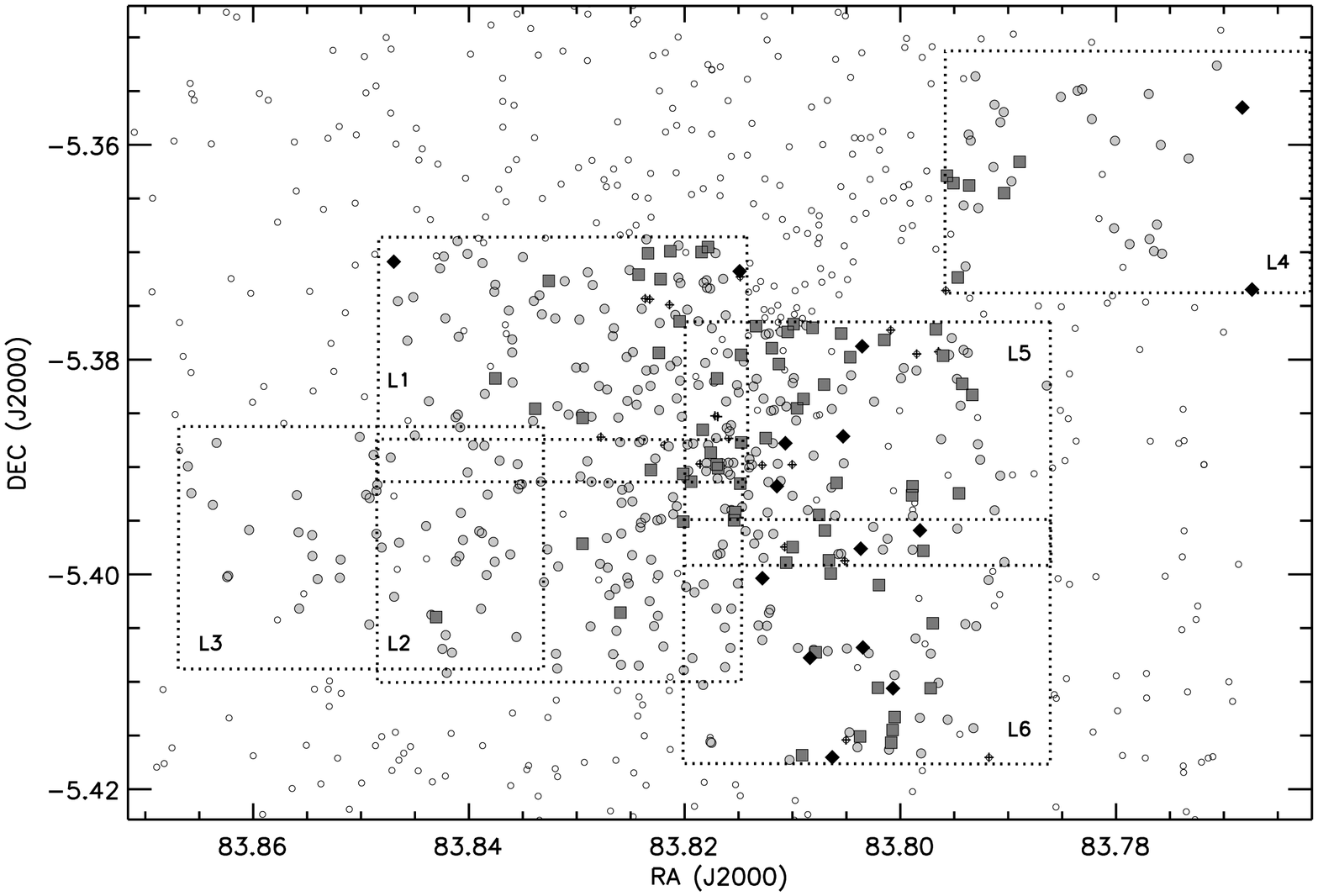}
\end{center}

\clearpage
\newpage
 \begin{center}
 {\bf Figure 2 is included as jpeg file in the ASTRO-PH version.}
 \end{center}

\clearpage
\newpage
\begin{center}
\includegraphics[angle=180,totalheight=8.0in]{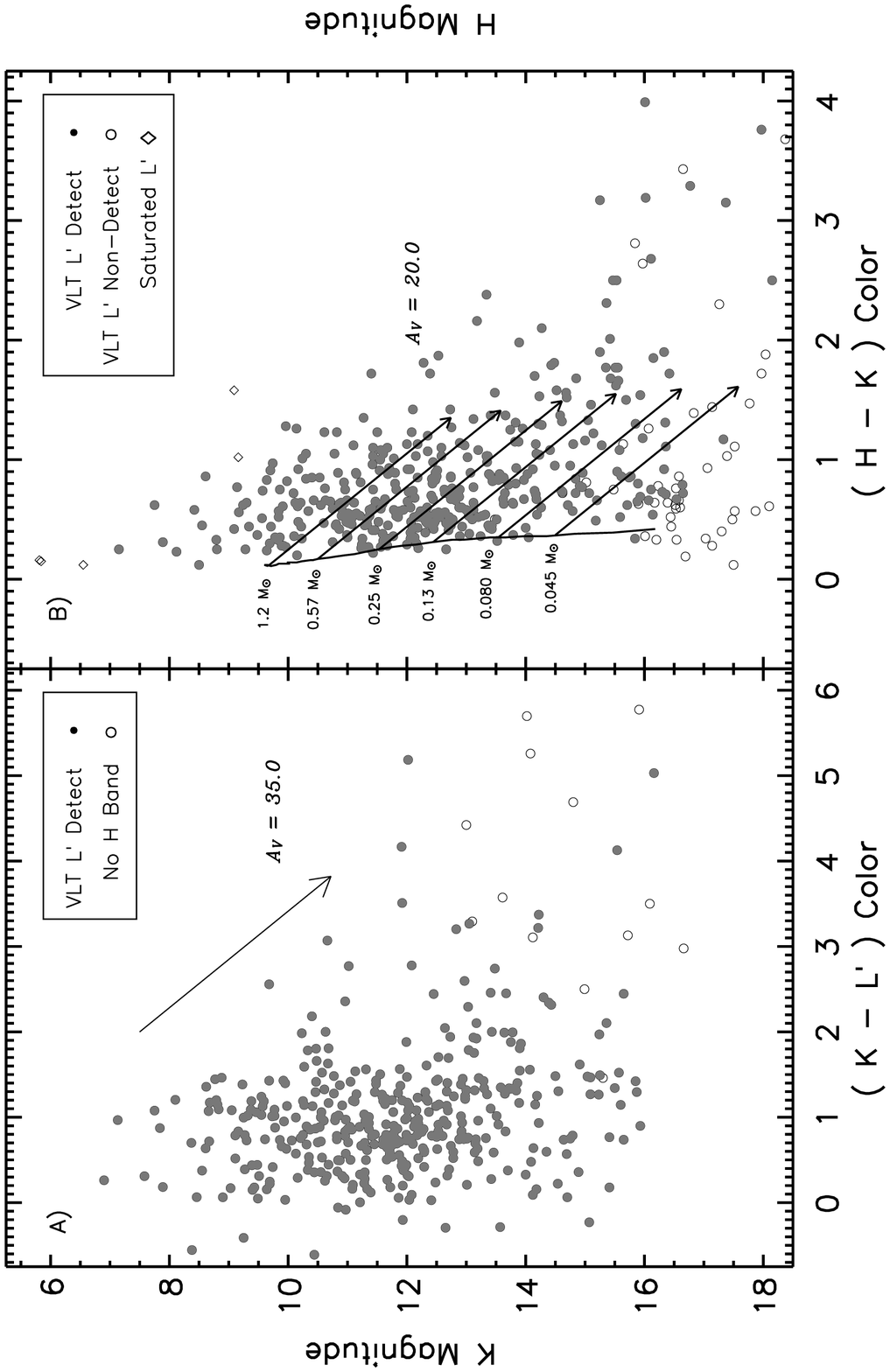}
\end{center}

\clearpage
\newpage
\begin{center}
\includegraphics[angle=180,totalheight=8.0in]{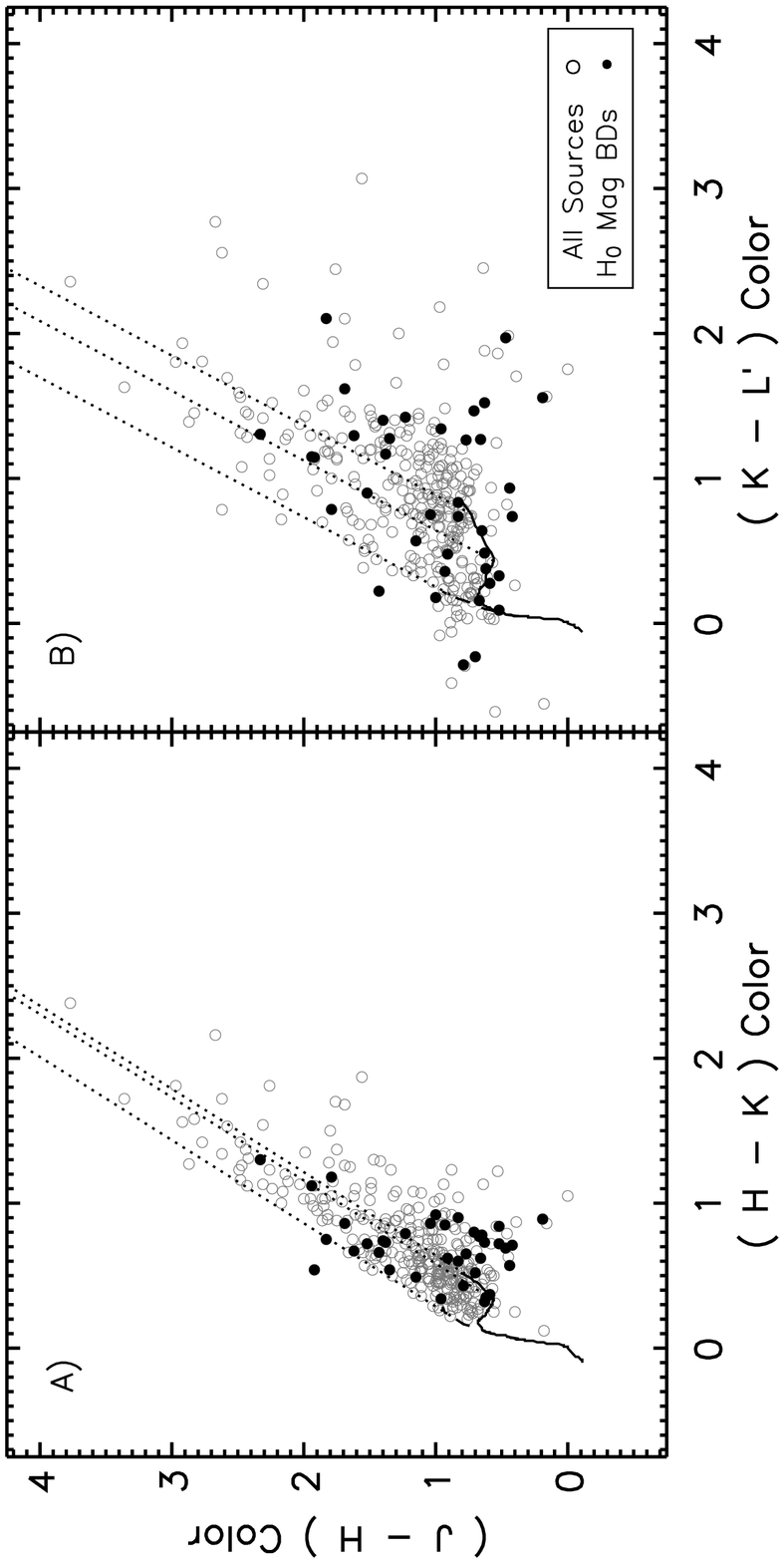}
\end{center}

\clearpage
\newpage
\begin{center}
\includegraphics[angle=180,totalheight=8.0in]{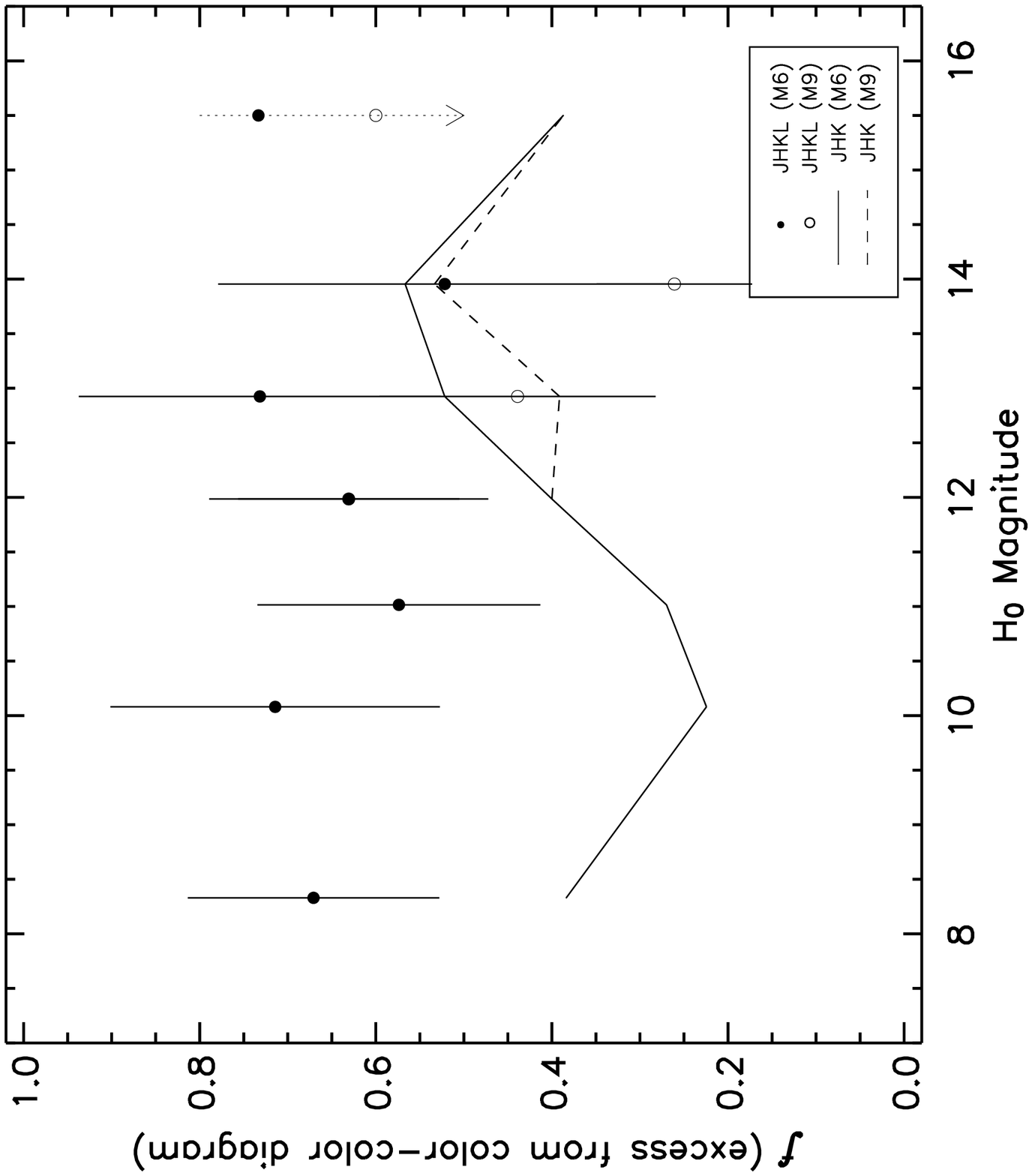}
\end{center}

\clearpage
\newpage
\begin{center}
\includegraphics[angle=180,totalheight=8.0in]{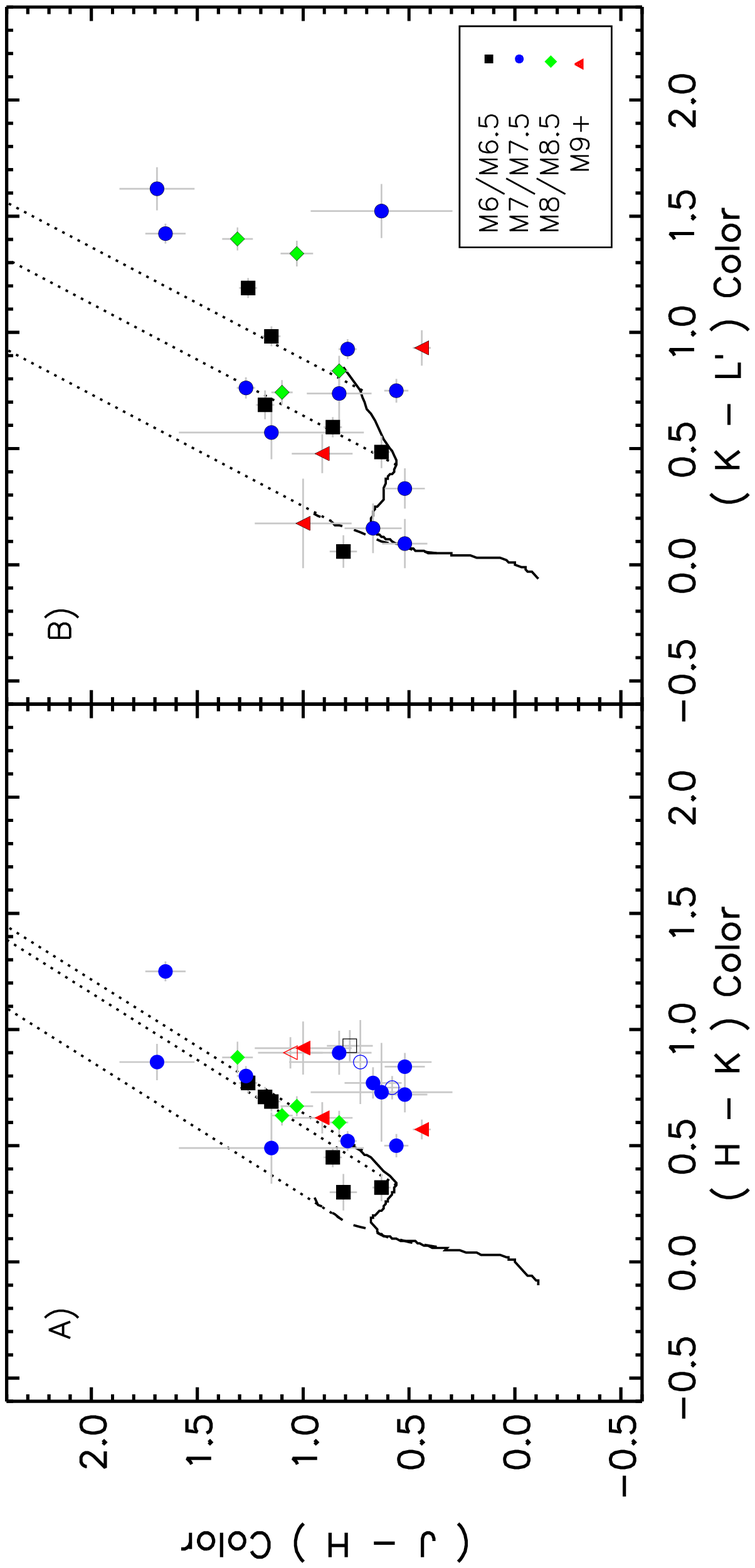}
\end{center}

\clearpage
\newpage
\begin{center}
\includegraphics[angle=180,totalheight=8.0in]{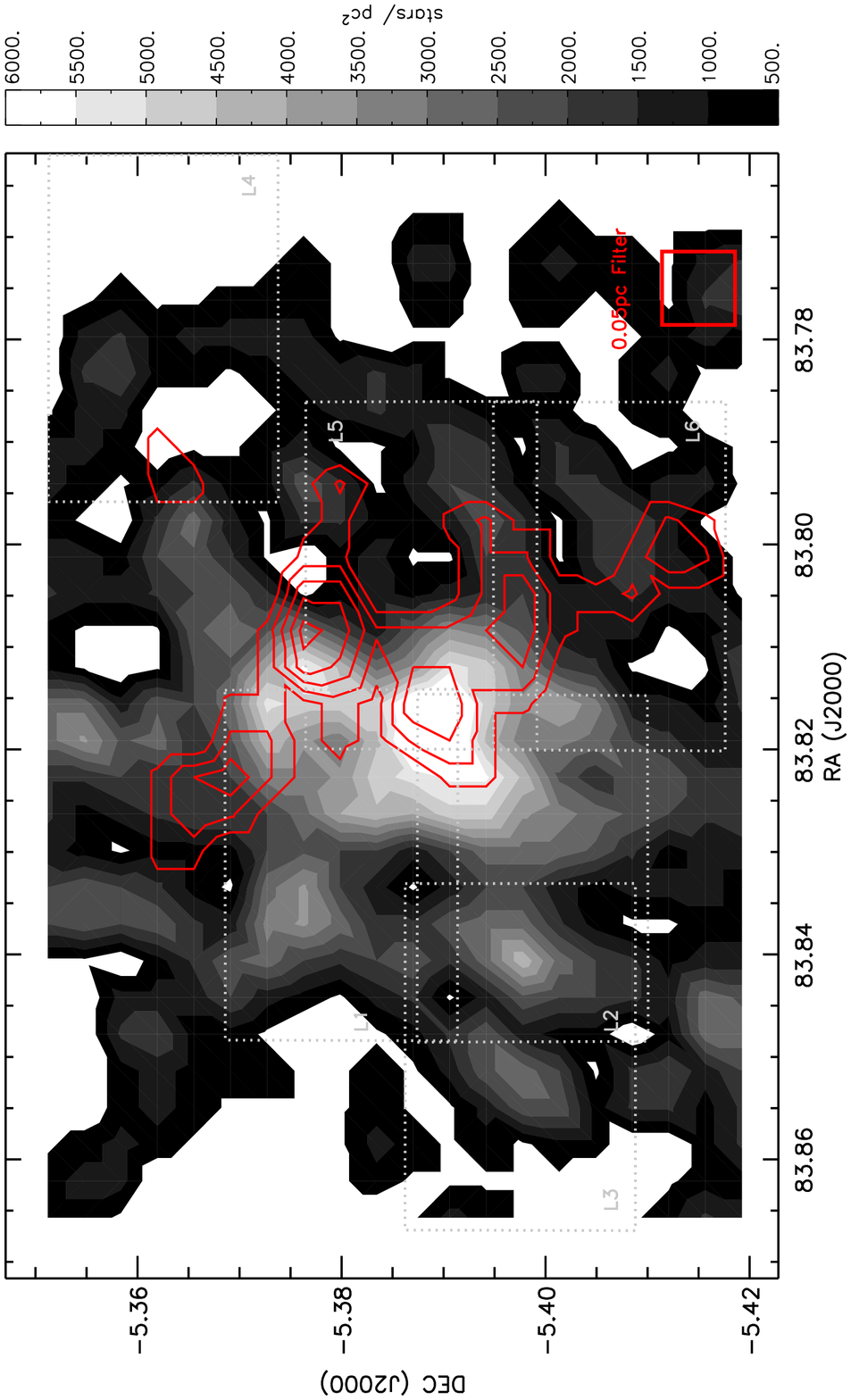}
\end{center}

\clearpage

%

\begin{deluxetable}{cccccccc}

\tablecaption{Observational Details\label{tab:fields}}
\tablehead{ 
\colhead{Image}         &
\colhead{Date}          &
\colhead{Julian Date}   &
\colhead{RA }           &
\colhead{DEC}           &
\colhead{Nods}          &
\colhead{Total Exp.}    &
\colhead{FWHM }         \\
\colhead{Name}          &
\colhead{}              &
\colhead{}              &
\colhead{(J2000)}       &
\colhead{(J2000)}       &
\colhead{Used}          &
\colhead{(min)\tablenotemark{(a)} } &
\colhead{($\arcsec$)}
}

\startdata
    L1  & 2002-10-10 & 52557.242325 & 5:35:19.507 & -5:22:47.95 & 60 & 29.70 & 0.53 \\
    L2  & 2002-10-10 & 52557.284110 & 5:35:19.580 & -5:23:55.45 & 59 & 29.21 & 0.34 \\
    L3  & 2002-10-12 & 52559.242141 & 5:35:24.003 & -5:23:51.06 & 59 & 29.21 & 0.85 \\
    L4  & 2002-10-12 & 52559.282716 & 5:35:06.943 & -5:21:45.10 & 60 & 29.70 & 0.39 \\
    L5  & 2002-10-14 & 52561.357352 & 5:35:12.723 & -5:23:16.17 & 59 & 29.21 & 0.48 \\
    L6  & 2002-12-19 & 52627.255228 & 5:35:12.749 & -5:24:22.58 & 60 & 29.21 & 0.42 \\
\enddata
\tablenotetext{(a)}{Total Exposure is derived from: Number of Nods * 2 (On+Off Chop) * 15 Chopping
cycles * 9 coadds * 0.11 sec per frame.}
\end{deluxetable}

\clearpage


\clearpage

\begin{deluxetable}{lrllrcrrcrrrrr}
\tablecolumns{14}
\tablewidth{0pt}

\tabletypesize{\tiny}
\tablecaption{VLT $3.8\micron$~Trapezium Catalog
\label{tab:catalog}}

\tablehead{
\colhead{No.}  &
\colhead{$f_n$} &
\colhead{R.A.}  &
\colhead{Dec.}  &
\colhead{$f_p$} &
\multicolumn{3}{c}{Image Information} &
\multicolumn{3}{c}{Photometry} &
\multicolumn{3}{c}{Cross References} \\

\colhead{}      &
\colhead{\tablenotemark{(a)}}      &
\colhead{(J2000)} &
\colhead{(J2000)} &
\colhead{\tablenotemark{(b)}}      &
\colhead{Image }&
\colhead{X Pixel}    &
\colhead{Y Pixel}    &
\colhead{Beam($\arcsec$)}    &
\colhead{$\Lp$}  &
\colhead{err}   &
\colhead{MLLA02} &
\colhead{HC2000} &
\colhead{H97}    
}
\startdata
      1 &  9 &05 35 04.174 &-05 22 24.52 & 0 &L4-033 &274.916& 1134.839& 1.14&  12.041&   0.056&        &      &     244 \\
      2 &  9 &05 35 04.389 &-05 21 23.51 & 0 &L4-012 &319.637&  268.436& 1.14&  14.007&   0.126&        &      &         \\
      3 &  1 &05 35 04.956 &-05 21 09.45 & 0 &L4-032 &439.672&   68.905& 1.14&  10.767&   0.019&   00856&      &         \\
      4 &  1 &05 35 05.583 &-05 21 40.58 & 0 &L4-011 &572.784&  510.949& 1.14&  14.428&   0.132&   00789&      &         \\
      5 &  1 &05 35 06.175 &-05 22 12.51 & 0 &L4-001 &698.199&  963.988& 1.14&  10.811&   0.019&   00681&   508&    3064 \\
      6 &  1 &05 35 06.203 &-05 21 36.06 & 0 &L4-013 &703.921&  446.854& 1.14&  15.031&   0.187&   00796&   601&         \\
      7 &  1 &05 35 06.289 &-05 22 02.75 & 0 &L4-004 &722.275&  825.489& 1.14&   8.593&   0.007&   00716&   538&     286 \\
      8 &  1 &05 35 06.355 &-05 22 11.63 & 0 &L4-002 &736.148&  951.438& 1.14&  14.575&   0.138&   00685&   509&         \\
      9 &  1 &05 35 06.454 &-05 22 07.61 & 0 &L4-003 &757.271&  894.435& 1.14&  11.691&   0.029&   00700&   526&     287 \\
     10 &  1 &05 35 06.472 &-05 21 18.97 & 0 &L4-009 &760.816&  204.533& 1.14&  11.755&   0.030&   00837&   636&     288 \\
     11 &  1 &05 35 06.896 &-05 22 09.31 & 0 &L4-005 &850.663&  918.522& 1.14&  12.715&   0.048&   00692&   521&     298 \\
     12 &  0 &05 35 7.046  &-05 22 17.06 & 0 &       &       &         &     &  99.000&  -1.000&   00667&   497&         \\
     13 &  1 &05 35 07.223 &-05 21 34.64 & 0 &L4-007 &919.862&  426.863& 1.14&  13.624&   0.077&   00803&   603&         \\
     14 &  1 &05 35 07.242 &-05 22 03.97 & 0 &L4-006 &923.860&  842.755& 1.14&  13.245&   0.063&   00711&   534&         \\
     15 &  0 &05 35 7.505  &-05 21 45.93 & 0 &       &       &         &     &  99.000&  -1.000&   00772&   718&         \\
\enddata

\tablenotetext{(a)}{Detection/Photometry Flag. The value of this flag indicates if a known source was saturated (-13), 
fell into a negative chop image (-1) or was undetected (0); it also indicates the number of times a detected source 
was photometered (1..3), in which case the photometry is the mean of the set and the recorded error is the standard
deviation.  Additional values of this flag are (9) for previously undetected IR sources and (10) for previously unresolved 
IR source.}
\tablenotetext{(b)}{Astrometry Flag. For sources falling into image overlaps, the relative dispersion of the astrometry
is recorded, otherwise the flag is set to 0. A flag of '1' indicates the relative dispersion was greater than $0.2\arcsec$; 
a value of '2' indicates the repeated source's astrometry was better than $0.2\arcsec$.}

\tablecomments{The complete version of this table is in the electronic
edition of the Journal.  The printed edition contains only a sample.}
\tablecomments{This table is available only on-line as a machine-readable table.}

\end{deluxetable}

\clearpage
%

\begin{deluxetable}{ccccccccccc}
\tablewidth{0pt}
\tabletypesize{\footnotesize}
\tablecaption{Excess Fractions for Trapezium Sources from Color-Color Diagrams \label{tab:trap_cc}}
\tablehead{ 
\colhead{Band}            & 
\multicolumn{2}{c}{$H_0$} & 
\multicolumn{2}{c}{$\solarmass$~\tablenotemark{(a)}} & 
\colhead{N$_{CMD}$} \tablenotemark{(b)}& 
\colhead{No $\Lp$} & 
\colhead{Boundary} & 
\colhead{$JHK$} & 
\colhead{$JHK\Lp$} & 
\colhead{$HK\Lp$} \\ 
\colhead{    }  & 
\colhead{Max}   & 
\colhead{Min}   & 
\colhead{Max}   & 
\colhead{Min}   & 
\colhead{(b)}   & 
\colhead{(c)}   & 
\colhead{(SpT)} & 
\colhead{$\%$}  & 
\colhead{$\%$}  & 
\colhead{$\%$}  } 

\startdata

1 &  7.00 &  9.66 & 5.000 & 1.200 & 83 &  0  & M6 & 38 & 67 & 66  \\
2 &  9.66 & 10.50 & 1.200 & 0.570 & 49 &  0  & M6 & 22 & 71 & 73  \\
3 & 10.50 & 11.53 & 0.570 & 0.250 & 63 &  0  & M6 & 27 & 57 & 60  \\
4 & 11.53 & 12.44 & 0.250 & 0.130 & 70 &  0  & M6 & 40 & 63 & 69  \\
5 & 12.44 & 13.41 & 0.130 & 0.080 & 43 &  1  & M6 & 52 & 73 & 76  \\
6 & 13.41 & 14.50 & 0.080 & 0.045 & 34 &  7  & M6 & 57 & 52 & 44  \\
7 & 14.50 & 16.50 & 0.045 & 0.020 & 45 &  28 & M6 & 39 & 73 & 71  \\

\tableline

5 & \nodata & \nodata & \nodata & \nodata & \nodata & \nodata & M9 & 39 & 44 & 43  \\
6 & \nodata & \nodata & \nodata & \nodata & \nodata & \nodata & M9 & 53 & 26 & 30  \\
7 & \nodata & \nodata & \nodata & \nodata & \nodata & \nodata & M9 & 39 & 60 & 65  \\

\tableline

\multicolumn{7}{l}{All $\Lp$~Sources} & M5 & 50 & 68 & 70 \\
\multicolumn{7}{l}{\nodata}           & M6 & 39 & 67 & 68 \\
\multicolumn{7}{l}{\nodata}           & M9 & 32 & 42 & 47 \\

\enddata

\tablenotetext{(a)}{Conversion from $H_0$ Magnitude to solarmass using the 1 Myr isochrone from \citet{Baraffe}}
\tablenotetext{(b)}{Total VLT sources with $HK$ photometry selected from 
Figure \ref{fig:vlt_cmds}b that are not saturated or falling into negative chop images.}
\tablenotetext{(c)}{Number of CMD selected sources undetected at $\Lp$.}

\end{deluxetable}

\clearpage
%

\begin{deluxetable}{lcccrrr}

\tablewidth{0pt}
\tablecaption{Sub-Clusters in the Trapezium\label{tab:subclusters}}
\tablehead{ 
\colhead{Sub-Cluster}   &
\colhead{RA }           &
\colhead{DEC}           &
\colhead{Peak Density}  &
\colhead{R\tablenotemark{(a)}} &
\colhead{N\tablenotemark{(b)}} &
\colhead{Comments}     \\
\colhead{}              &
\colhead{(J2000)}       &
\colhead{(J2000)}       &
\colhead{stars/pc-2}    &
\colhead{(pc)}          &
\colhead{}              &
\colhead{}
}

\startdata
%
%
\multicolumn{7}{c}{Trapezium Sub-Clusters\tablenotemark{(c)}} \\
  TSC-a &     05:35:16.234 &     -05:23:25.19 &   7200. &    0.08 & 93 & Trapezium Core \\
  TSC-b &     05:35:15.734 &     -05:22:29.79 &   5600. &    0.06 & 43 & Near BNKL \\
  TSC-c &     05:35:21.689 &     -05:23:51.70 &   4400. &   <0.05 & 16 & \\
  TSC-d &     05:35:20.490 &     -05:22:33.72 &   3600. &    0.05 & 20 & \\
%
%
\multicolumn{7}{c}{Deeply Embedded $K-L$~ Sub-Clusters\tablenotemark{(d)}} \\
  DESC-a &     05:35:15.672 &     -05:23:24.92 &   2400. &    0.05 & 14 & Trapezium Core \\
  DESC-b &     05:35:14.090 &     -05:22:41.14 &   3200. &    0.06 & 16 & BNKL \\
  DESC-c &     05:35:17.739 &     -05:22:05.22 &   2400. &    0.05 & 12 & \\
  DESC-d &     05:35:13.989 &     -05:23:49.94 &   2400. &   <0.05 & 10 & Orion S.\\
  DESC-e &     05:35:12.264 &     -05:24:43.50 &   2000. &   <0.05 &  8 & \\
 
\enddata
\tablenotetext{(a)}{Sub-Cluster radius suggested by radial profile of candidates sources (low $\av$ or embedded).}
\tablenotetext{(b)}{Number of sub-cluster members within radius.}
\tablenotetext{(c)}{Trapezium sources defined by $J-H < 1.5$, which for a CTTS corresponds to approximately $\av\sim8$.}
\tablenotetext{(d)}{Embedded sources defined by $K-L<1.5$ and $L$ only sources.}
\end{deluxetable}

\end{document}